\documentclass{ifacconf}

\usepackage{amsmath,amssymb}
\usepackage{graphicx}          % Include this line if your 
\usepackage{color}
\usepackage{setspace}

\newtheorem{defi}{Definition}
\newtheorem{assume}{Assumption}
\newtheorem{exx}{Example}
\newtheorem{remm}{Remark}

\newenvironment{theorem}{\begin{thm}\rm }
{\hfill \hspace*{1pt} \hfill $\lrcorner$ \end{thm}}

\newenvironment{definition}{\begin{defi}\rm }
{\hfill \hspace*{1pt} \hfill $\lrcorner$ \end{defi}}
\newenvironment{remark}{\begin{remm}\rm }
{\hfill \hspace*{1pt} \hfill $\lrcorner$\end{remm}}
\newenvironment{example}{\begin{exx}\rm }
{\hfill \hspace*{1pt} \hfill $\lrcorner$ \end{exx}}
\newenvironment{proofof}{\noindent {\em Proof of }}{\hfill \hspace*{1pt}
\hfill $\square$}

\newcommand\real{\ensuremath{{\mathbb R}}}
\newcommand\realn{\ensuremath{{\mathbb{R}^n}}}

\newcommand{\calM}{\mathcal{M}}
\newcommand{\calU}{\mathcal{U}}
\newcommand{\calY}{\mathcal{Y}}
\newcommand{\calW}{\mathcal{W}}
\newcommand{\bbW}{\mathbb{W}}

\newcommand{\smallmat}[1]{\left[ \begin{smallmatrix} #1 \end{smallmatrix} \right]}

\newcommand{\inner}[1]{ \langle #1 \rangle }
\begin{document}

\begin{frontmatter}

\title{On differentially~dissipative dynamical~systems} % Title, preferably not more than 10 words.

\thanks[footnoteinfo]{This paper presents research results of the Belgian Network DYSCO 
(Dynamical Systems, Control, and Optimization), funded by the 
Interuniversity Attraction Poles Programme, initiated by the Belgian State, 
Science Policy Office. The scientific responsibility rests with its author(s).
F. Forni is supported by FNRS.}

\author[authors]{F. Forni}, 
\author[authors]{R. Sepulchre} 

\address[authors]{Systems and Modeling,  
Department of Electrical Engineering\\ and Computer Science,
University of Li{\`e}ge, Belgium; fforni$|$r.sepulchre@ulg.ac.be.}

                             % keyword list or with the 
                                          % help of the Automatica 
                                          % keyword wizard

\begin{abstract}                          % Abstract of not more than 250 words.
Dissipativity is an essential concept of systems theory. 
The paper provides an extension of dissipativity, named differential dissipativity,
by lifting storage functions and supply rates to the tangent bundle. 
Differential dissipativity is connected to incremental stability in the same way
as dissipativity is connected to stability. 
It leads to a natural formulation of differential passivity when restricting 
to quadratic supply rates. The paper also shows
that the interconnection of differentially passive systems is differentially passive,
and provides preliminary examples of differentially passive electrical systems. \vspace{-2mm}
\end{abstract}

\end{frontmatter}

\section{Introduction}
\label{sec:introduction}
Dissipativity, \cite{Willems1972}; \cite{Willems1972a},
plays a central role in the analysis of open systems
to reduce the analysis of complex systems to the
study of the interconnection of simpler components.
Dissipativity is a fundamental tool in nonlinear control design 
\cite{Sepulchre1997,VanDerSchaft1999}, widely 
adopted in industrial applications. Typical examples are
provided by applications on electro-mechanical devices 
modeled within the port-hamiltonian framework, \cite{Ortega2001}. 
Passivity-based designs conveniently connect the physical modeling
of mechanical and electrical interconnections and the stability
properties required by applications.

In a nonlinear setting, applications like regulation, observer designs, and synchronization 
call for  incremental notions of stability, 
 \cite{Angeli00}; \cite{Angeli09}.
Several results in the literature propose extensions of passivity to
guarantee connections to incremental properties. For example, 
in the theory
of equilibrium independent passivity, \cite{Hines2011,Jayawardhana2007}, 
the dissipation inequality refers to pairs of system trajectories, 
one of which is a fixed point. 
The incremental passivity of \cite{Desoer1975} and \cite{Stan2007} characterizes a 
passivity property of solutions pairs,  through the use of 
incremental storage functions reminiscent of the
notion of incremental Lyapunov functions of \cite{Angeli00}, and
supply rates of the form $Q:= \Delta y^T \Delta u$, for
$\Delta y := y_1-y_2$ and $\Delta u := u_1-u_2$, where $u_i$ and $y_i$ 
refers to input/output signals.

Incremental passivity is equivalent to passivity for linear systems. 
It has been used in nonlinear control for regulation, \cite{Pavlov2008},
and synchronization purposes, \cite{Stan2007}. Yet, 
it requires the construction of a storage function in 
the extended space of paired solutions, a difficult task in general,
and the a priori formulation of the supply rate based on the difference between
signals, which does not take into account the possible nonlinearities of the 
state and external spaces.
A motivation for the present work partly come A motivation for the present paper partly comes from the
role of incremental properties in ant windup design of induction motors \cite{Sepulchre11}
and the difficulty to establish those properties in models that integrate
magnetic saturation, see Example \ref{ex:induction_motor} in the present paper.

A different approach to the characterization of incremental properties
is provided by contraction, a differential concept
The theory developed in \cite{Lohmiller1998}
recognizes that the infinitesimal approximation of a system
carries information about the behavior of its solutions set. 
It provides a variational approach to  incremental stability, 
based on the linearization of the system, 
without explicitly constructing the distance 
measuring the convergence of solutions towards each other. 

Following this basic idea, 
the present paper proposes a dissipativity theory 
based on the infinitesimal variations of dynamical systems along their solutions. 
We call it \emph{differential dissipativity} because it is
classical dissipativity lifted to  the tangent bundle of the
system manifold. 
In analogy with the classical relation 
between storage functions and Lyapunov functions,
the proposed notion of \emph{differential storage function} for differential dissipativity 
is paired to the notion of Finsler-Lyapunov function recently proposed in \cite{Forni2012}, 
which plays a role in connecting differential dissipativity and incremental stability. 
The preprint \cite{Schaft2013submitted} is an insightful complementary effort in that direction, 
connecting the framework to the early concept of prolonged system
in nonlinear control \cite{crouch1987}.

The are many potential advantages in developing a differential version of dissipativity theory. 
First of all, differential dissipativity is equivalent to dissipativity for linear systems.
In the nonlinear setting, the fact that the infinitesimal approximation of a nonlinear system is a 
linear time-varying system opens the way to a characterization of 
\emph{differential passivity} - differential dissipativity with quadratic supply rates -
that falls in the linear setting of  
\cite{Willems1972a}. Moreover, 
differential dissipativity provides an input-output characterization of the dynamical 
system in the infinitesimal neighborhood of each trajectory, which leads to
state-dependent differential supply rates. This is of relevance to tailor the
dissipativity property to nonlinear state and external variables spaces.

The content of the paper is developed in analogy with classical results on dissipativity.
The instrumental notion of displacement dynamical system is provided in 
Section \ref{sec:dds}. Differential dissipativity and differential passivity 
are formulated in Sections \ref{sec:dd}  and  \ref{sec:dp}, 
Examples of differentially passive 
electromechanical systems are proposed in Section \ref{sec:ems}. Conclusion follows. 
Proofs are in appendix. This paper is an extended version of \cite{Forni2013}.

{\small
\textbf{Notation.}
The exposition of the differential dissipativity approach is developed on manifolds
following the notation of \cite{AbsMahSep2008} and \cite{DoCarmo1992}.

Given a manifold $\calM$, and a point $x$ of $\calM$, 
$T_x\calM$ denotes the \emph{tangent space} of $\calM$ at $x$.
$T\calM:= \bigcup_{x\in\mathcal{M}} \{x\}\times T_x\calM$
is the \emph{tangent bundle}.
Given two manifolds $\calM_1$ and $\calM_2$ and a mapping 
$F:\calM_1 \to \calM_2$. $F$ is
of class $C^k$, $k\in\mathbb{N}$,  if the function 
$\hat{F} = \varphi_2 \circ F \circ \varphi_1^{-1}:\real^{d_1}\to\real^{d_2}$ is of class $C^k$,
where $\varphi_1:\mathcal{U}_x\subset\calM_1\to\real^{d_1}$ and 
$\varphi_2:\mathcal{U}_{F(x)}\subset\calM_2\to\real^{d_2}$ are smooth charts.
The \emph{differential of $F$ at $x$} is denoted by 
$DF(x)[\cdot]:T_x \calM_1 \!\to\! T_{F(x)}\calM_2$.
A \emph{curve} $\gamma$ on a given manifold $\calM$ is a mapping 
$\gamma :I \subset \real \to \calM$.
For simplicity we sometime use
$\dot{\gamma}(t)$ or $\frac{d\gamma(t)}{dt}$ to denote $D\gamma(t)[1]$.
Specifically, this notation is adopted when the variable $t$ in $\gamma$ refers to time.

$I_n$ is the identity matrix of dimension $n$.
Given a vector $v$, $v^T$ denotes the transpose vector 
of $v$. Given a matrix $M$ we say that 
$M\geq 0$ or $M\leq 0$  if $v^T M v \geq 0 $ or $v^T M v \leq 0$,
for each $v$, respectively.
Given the vectors $\{v_1,\dots,v_n\}$, 
$\mathrm{Span}(\{v_1,\dots,v_n\}) := \{v\,|\,\exists \lambda_1,\dots\lambda_n\in\real \mbox{ s.t. } v = \sum_{i=1}^n \lambda_i v_i\}$.
A locally Lipschitz function 
$\alpha:\real_{\geq 0}\rightarrow\real_{\geq 0}$ is said 
to belong to \emph{class} $\mathcal{K}$ if it is strictly increasing and $\alpha(0) = 0$;
it belongs to \emph{class} $\mathcal{K}_{\infty}$ if, moreover,
 $\lim_{r \rightarrow+\infty}\alpha(r)=+\infty$.

A \emph{distance} (or \emph{metric}) $d:\calM\times\calM\to \real_{\geq 0}$ 
on a manifold $\calM$ is a positive function that satisfies
$d(x,y) = 0$ if and only if $x=y$, for each $x,y\in \calM$ and 
$d(x,z) \leq d(x,y) + d(y,z)$ for each $x,y,z\in \calM$. 
If $d(x,y) = 0$ but $x\neq y$ we say that $d$ is a pseudo-metric.
A set $\mathcal{S}\subset\mathcal{M}$ is bounded if $\sup_{x,y\in\mathcal{S}} d(x,y) < \infty$ for any given
distance $d$ on $\mathcal{M}$. A curve $\gamma:I\to\calM$ is \emph{bounded} when its image is bounded.
Given a manifold $\calM$, a set of \emph{isolated points} $\Omega\subset \calM$
satisfies: for any distance function $d$ on $\calM$ and any given pair 
$x_1,x_2$ in $\Omega$, there exists an $\varepsilon>0$ such that $d(x_1,x_2)\geq \varepsilon$.
Given $f:\mathcal{Z}\to\mathcal{Y}$ and $g:\mathcal{X}\to\mathcal{Z}$, 
the \emph{composition} $f\circ g$ assigns to each  
$p\in\mathcal{X}$ the value $f\!\circ\! g(p) = f(g(p)) \in \mathcal{Y}$.
Given a function $f:\real^n\to\real^m$, the matrix of partial derivatives is denoted as
$\partial_x f(x)$ (Jacobian).  $\partial_{xx}f(x)$ denotes the Hessian of $f(x)$.
}

\section{Displacement dynamical systems}
\label{sec:dds}
Taking inspiration from the dissipativity paper of
\cite{Willems1972} and from the (state-space) behavioral framework in
\cite{Willems1991},  given 
smooth manifolds $\calM$ and $\calW$,
a time-invariant dynamical system $\Sigma$
is represented by algebraic-differential equations of the form
\begin{equation}
\label{eq:dynsys}
 F(x,\dot{x},w) = 0 \ ,
\end{equation}
where $F:T\calM\times\calW\to\real^p$, $p\in\mathbb{N}$,
$x\in\mathcal{M}$ is the state, 
and $w$ collects the external variables.
The behavior of $\Sigma$ is given by the set of 
{ absolutely continuous} curves $(x,w)(\cdot):\real \to \calM\times\calW$ 
that satisfy $F(x(t),Dx(t)[1],w(t))=0$ for (almost) all $t\in\real$.
Given $w = (u,y)$, $u$ - input, $y$ - output, and 
$(x,u,y)(\cdot)\in\Sigma$, we say that 
$x(\cdot)$ is a \emph{solution} to \eqref{eq:dynsys} from the
initial condition $x(0)\in\calM$ under the action of the input $u(\cdot)$.

In what follows we assume that $(x,w)(\cdot)\in\Sigma$ are $C^2$
functions. {
When the external variables are organized into input
and output variables, i.e. $w =(u,y)$
}, we also assume existence, unicity, and forward completeness
of solutions for each initial condition $x_0$ and input $u(\cdot)$. 
Note that under mild regularity assumptions on $F$, if 
$u(\cdot)\in C^2$, every
$(x,u,y)(\cdot)  \in \Sigma$ is a $C^2$ curve,
as clarified in Chapter IV, Section 4, of \cite{boothby2003}.

Under these assumptions, the 
\emph{displacement dynamical system} $\delta \Sigma$ induced by
$\Sigma$ is represented by 
\begin{subequations}
\label{eq:ddynsys}
\begin{eqnarray}
F(x,\dot x, w) &=& 0 \\
 DF(x,\dot{x},w)[\delta x,\dot{\delta x}, \delta w] &=& 0 \ ,
 \end{eqnarray}
\end{subequations} 
and it is given by the set of $C^1$ curves
$(x,\delta x,w,\delta w)(\cdot):\real \to 
T\calM \times T\calW $ that satisfy
\eqref{eq:ddynsys} for each $t\in \real$.

\begin{figure}[htbp]
\centering
\includegraphics[width=0.6\columnwidth]{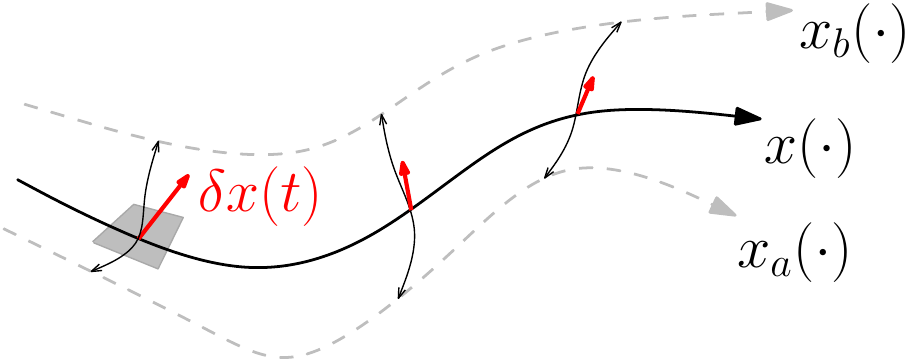}  
\caption{The tangent vector $\delta x(t)$ represents an infinitesimal
variation on $x(t)$. 
Given an input curve $u(\cdot)$ and its infinitesimal 
variation $\delta u(\cdot)$, the time-evolution of $\delta x(\cdot)$ along 
a given solution $x(\cdot)$ to \eqref{eq:dynsys} must satisfy \eqref{eq:ddynsys}. 
A precise characterization is given in Remark \ref{rem:d_and_(2)}.
} 
\label{fig:displacement}
\end{figure}

Following the interpretation proposed in \cite{Lohmiller1998}, 
given a point $(x,w)\in \calM\times\calW$,
a tangent vector 
$(\delta x,\delta w)\in T_x\calM \times T_w\calW $
represents an infinitesimal variation - or displacement - on $(x,w)$. 
In this sense $\delta \Sigma$ characterizes the infinitesimal
difference between every two neighborhood solutions, that is,
the infinitesimal variations $\delta x(\cdot)$
on the solutions $x(\cdot)$ to \eqref{eq:dynsys}. 
A graphical representation of a displacement is proposed
in Figure \ref{fig:displacement}. The intuitive notion of
infinitesimal variation  is made precise in Remark \ref{rem:d_and_(2)}.
\begin{remark}
\label{rem:d_and_(2)}
For each $s\in[0,1]$, 
consider a (parameterized) curve $(x,w)(\cdot,s):\real\to\calM\times\calW \in \Sigma$.
We assume that $(x,w)(\cdot,\cdot) \in C^2$. 
An infinitesimal variation on $(x,w)(\cdot,s)$ is given by
$(\delta x,\delta w)(\cdot,s) := (Dx(\cdot,s)[0,1],Dw(\cdot,s)[0,1])$. 
As a matter of fact, 
$(x, \delta x,w,\delta w)(\cdot,s)\in \delta \Sigma$ for each $s\in[0,1]$.
In fact, by chain rule\footnote{The differential in the right-hand side
of the first identity refers to the mapping from $\real\times [0,1]$ to
$\real^p$. The one in the right-hand side of the second identity refers 
to the mapping from
$T \calM \times \calW$ to $\real^p$.}, 
\begin{equation}
\begin{array}{rcl}
 0 
 &=& DF(x(t,s),\dot{x}(t,s),w(t,s))[0,1]  \vspace{1mm}\\
 &=&  DF(\dots)[\delta x(t,s), D\dot{x}(t,s)[0,1], \delta w(t,s)] \vspace{1mm}\\
&=& DF(\dots)[\delta x(t,s), \dot{\delta x}(t,s), \delta w(t,s)]
 \end{array}
\end{equation}
where the third identity follows from the fact that 
$x(\cdot,\cdot)$ is a $C^2$ function, by assumption (in local coordinates
$\partial_s\partial_t x(t,s) = \partial_t\partial_s x(t,s)$.
\end{remark}
When the manifold $\calM$ is equipped with a { Finsler metric 
$|\delta x|_x$ (see, for example, \cite{Tamassy08,Shen00}),
the time-evolution of 
$|\delta x(t)|_{x(t)}$} along the solutions 
$(x(\cdot),\delta x(\cdot))$ to \eqref{eq:ddynsys}
measures the contraction of the dynamical system
$\Sigma$, that is, the tendency of solutions to converge
towards each other. The connection between the displacement
dynamical system $\delta\Sigma$ 
and incremental stability properties have been exploited in the
seminal paper of \cite{Lohmiller1998}, and
in many other works, e.g. \cite{Lewis1949,Rouchon2003,Pavlov04,Wang05,Fromion2005,Pham2007,Russo10}.
A unifying framework for contraction based on the extension of Lyapunov theory to
the tangent bundle has been recently proposed in \cite{Forni2012}.

\section{Differentially dissipative systems}
\label{sec:dd}
We develop the theory of differential dissipativity mimicking classical 
dissipativity, \cite{Willems1972,Sepulchre1997,VanDerSchaft1999}.
In analogy to the intuitive interpretation of a storage function
as the energy of the system, it is convenient to view the
\emph{differential storage function} $S:T\calM\to\real_{\geq 0}$ 
as the infinitesimal energy 
associated to the infinitesimal variation $\delta x(\cdot)$ on a given solution
$x(\cdot)$.
This energy
can be either increased or decreased through the 
supply provided by external sources, as prescribed by 
a \emph{differential supply rate}  $Q$. 
{
\begin{definition}
\label{def:storage}
 Consider a manifold $\mathcal{M}$ and a set of isolated points $\Omega\subset\mathcal{M}$. 
 For each $x\in\mathcal{M}$, consider a subdivision of $T_x\mathcal{M}$ into a \emph{vertical distribution}
$\mathcal{V}_x\subset T_x\mathcal{M}$ 
 \begin{equation}
  \label{eq:V_x}
  \mathcal{V}_x = \mathrm{Span}(\{v_1(x),\dots, v_r(x)\}), \quad 0\leq r<d \ ,
 \end{equation}
 and a \emph{horizontal distribution} $\mathcal{H}_x\subseteq T_x\mathcal{M}$ 
 complementary to $\mathcal{V}_x$, i.e.
 $\mathcal{V}_x \oplus \mathcal{H}_x = T_x\mathcal{M}$, given by  
 \begin{equation}
 \label{eq:H_x}
 \mathcal{H}_x = \mathrm{Span}(\{h_1(x),\dots, h_q(x)\}), \quad 0 < q \leq d-r
 \end{equation}
 { where $v_i$, $i \in \{1,\dots,r\}$, 
 and $h_i$, $i\in\{1,\dots,q\}$, are $C^1$ vector fields.}
 
 A function $S: T\mathcal{M} \to \real_{\geq 0}$  
 is a \emph{differential storage function}  for the dynamical system $\Sigma$ in \eqref{eq:dynsys} if 
 there exist $c_1,c_2\in \real_{\geq 0}$, $p\in\real_{\geq 1}$, 
 and  $K:T\mathcal{M} \to \real_{\geq 0}$ such that
 \begin{equation}
 \label{eq:FinsLyap_bounds}
  c_1\, K(x,\delta x)^p  \leq  S(x,\delta x)  \leq  c_2\, K(x,\delta x)^p  
 \end{equation}
 for all $(x,\delta x) \in T\mathcal{M}$,
 where $S$ and $K$ satisfies the  following conditions: 
 \begin{itemize}
  \item[(i)] $S$ and $K$ are $C^1$ functions for each $x\in\mathcal{M}$ and $\delta x\in \mathcal{H}_x \setminus\{0\}$;
  \item[(ii)] $S$ and $K$ satisfy 
  $S(x,\delta x) = S(x,\delta x_h)$ and $K(x,\delta x) = K(x,\delta x_h)$
  for each $(x,\delta x)\in T\mathcal{M}$ such that 
  $(x,\delta x) = (x,\delta x_h) + (x,\delta x_v)$, 
  $\delta x_h \in \mathcal{H}_x$, and $\delta x_v \in \mathcal{V}_x$. 
  \item[(iii)] $ K(x,\delta x) >0$ for each $x\in\mathcal{M}\setminus\Omega$ 
  and $\delta x\in \mathcal{H}_x\setminus\{0\}$.  
  \item[(iv)] $K(x,\lambda \delta x) = \lambda S(x,\delta x)$ 
  for each $\lambda>0$, $x\in\mathcal{M}$, 
   and $\delta x\in \mathcal{H}_x$; 
  \item[(v)] $ K(x,\delta x_1+\delta x_2) <  K(x,\delta x_1) +  K(x,\delta x_2)$ for each $x\in \mathcal{M}\setminus\Omega$ and 
        $\delta x_1,\delta x_2\in \mathcal{H}_x\setminus\{0\}$ such that $\delta x_1\neq \lambda \delta x_2$ and $\lambda \in \real$ (strict convexity).
 \end{itemize} \vspace{-6mm}
\end{definition}
\begin{definition}
A function $Q:\calM\times T\calW \to \real$ 
is a \emph{differential supply rate} for the dynamical system $\Sigma$ in \eqref{eq:dynsys}
if 
\begin{equation}
 \int_0^t | Q( x(\tau), w(\tau),\delta w(\tau)| d\tau < \infty  
\end{equation}
for each $t \geq 0$ and each $(x,\delta x,w,\delta w)(\cdot) \in \delta \Sigma$.
\end{definition}
}
The function $S$ provides a non-negative value $S(x,\delta x)$ to
each $\delta x \in T_x\calM$. When $\mathcal{V}_x=\emptyset$, 
{ a suggestive notation
for $K(x,\delta x)$ is $|\delta x|_x$ - a non-symmetric 
norm on each tangent space $T_x\mathcal{M}$ - which immediately connects
the differential storage to the idea 
of an energy of the displacement $\delta x$, since
$c_1 |\delta x|_x^p \leq S(x,\delta x) \leq c_2|\delta x|_x^p$. }
From Definition \ref{def:storage} it is possible to identify differential storage functions $S$
and horizontal Finsler-Lyapunov functions $V$, introduced 
in Section VIII of \cite{Forni2012}. Therefore the existence of a differential storage $S$
endows $\calM$ with the structure of a pseudo-metric space, which plays a central role 
in connecting differential dissipativity to incremental stability. 
Restricting a differential storage to horizontal distributions is convenient
in many situations where contraction takes place only in certain directions.
For example, let $\calM$ be the state space and suppose that the output 
$y\in\calY$ is given by $y= h(x)$ where $h:\calM\to\calY$ is a differentiable function. 
Then, in coordinates, $\delta y^T \delta y$ is a possible candidate storage function 
with horizontal distribution 
$\mathcal{H}_x$ given by the span of the columns of
the matrix $\partial_x h(x)^T \partial_x h(x) $. With this storage, 
the state-space $\calM$ becomes a pseudo-metric space, while the output space $\calY$
becomes a metric space. 
Further details are collected in Remark \ref{rem:metric_space}.

\begin{remark}
\label{rem:metric_space}
Suppose that for each $x\in\calM$, $\mathcal{H}_x = T_x \calM$,
and take $\Omega = \emptyset$. Then, $K$ 
is a Finsler structure on $\mathcal{M}$ (see, for example, \cite{Tamassy08,Shen00}).
Then, we can define the length of a curve as 
$ L(\gamma) := \int_I K(\gamma(s), D\gamma(s)[1]) ds$.
The induced distance $d$ between any two points $x_0,x_1\in\mathcal{M}$
is given by
$d(x_0,x_1) := \inf_{\Gamma(x_0,x_1)} L(\gamma)$,
where $\Gamma(x_0,x_1)$ is
the set of piecewise $C^1$ curves in $[0,1] \to \calM$ 
such that $\gamma(0) = x_0$ and $\gamma(1) = x_1$. 
For the case $\mathcal{H}_x \neq T_x\mathcal{M}$, we have the identity
$L(\gamma) = \int_I K(\gamma(s), \Pi_h(D\gamma(s)[1])) ds $,
where the function $\Pi_\mathcal{H}(\cdot)$ projects every tangent vector
$v_x\in T_x\calM$ into $\Pi_\mathcal{H}(v_x)\in \mathcal{H}_x$. In this case,
$L(\gamma)$ measures only the horizontal contribution of $\gamma$,
and the induced $d(x_0,x_1) := \inf_{\Gamma(x_0,x_1)} L(\gamma)$, is only 
a pseudo-distance on $\mathcal{M}$, since
$d(x_0,x_1) = 0$ for some $x_0\neq x_1$. 
An extended discussion and examples are provided in 
Sections IV and VIII of \cite{Forni2012}. 
\end{remark}

We can finally provide the definition of differential dissipativity. 
We emphasize that differential dissipativity is
just dissipativity lifted to the tangent bundle.
\begin{definition}
\label{def:dds}
The dynamical system $\Sigma$ in \eqref{eq:dynsys} is 
\emph{differentially
dissipative} with respect to the differential supply rate $Q$
if there exists a differential storage function $S$ such that 
\begin{equation}
\label{eq:dissipativity}
S( x(t),\delta x(t)) - S( x(0),\delta x(0)) 
\leq \int_0^t \!\! Q( x(\tau), w(\tau),\delta w(\tau)) d\tau 
\end{equation}
for all $t \geq 0$ and all $(x,\delta x,w,\delta w)(\cdot) \in \delta \Sigma$ in \eqref{eq:ddynsys}.
When $Q$ is independent on $x$,
that is, $Q:T\calW \to \real$, we say that 
$\Sigma$ is \emph{uniformly differentially dissipative}.
\end{definition}
{
Exploiting the assumption $S\in C^1$, \eqref{eq:dissipativity} 
is equivalent to}
\begin{equation}
\label{eq:diff_dissipativity}
 \frac{d}{dt} S(x(t),\delta x(t)) \leq Q( x(t), w(t),\delta w(t)).
\end{equation}

We conclude the section by illustrating a first connection 
between differential dissipativity and incremental stability. 
\begin{theorem}
\label{thm:dd2gs}
Suppose that the dynamical system $\Sigma$ represented 
by \eqref{eq:dynsys} is differentially dissipative
with differential storage $S$ and differential supply rate $Q$.
Suppose also that for $w = (u,y)$, $u$ - input, $y$ - output,
it holds that 
$Q(x,u,y,0,\delta y) = 0$ for each $x\in \calM$,
and each $(u,y,0,\delta y)\in T\calW$.
Then, 
there exists a class $\mathcal{K}$ function $\alpha$
such that 
\begin{equation}
\label{eq:dd2gs}
 d(x_1(t),x_2(t)) \leq  \alpha(d(x_1(0),x_2(0))) 
\end{equation}
for each $t\geq 0$ and each $(x_1,u_1,y_1)(\cdot),(x_2,u_2,y_2)(\cdot) \in \Sigma$,
such that $u_1(\cdot)=u_2(\cdot)$,
where $d$ is the pseudo-distance induced by $S^{\frac{1}{p}}$, with
$p$ degree of homogeneity of $S$ (see Definition \ref{def:storage}).
\end{theorem}
Note that if
$\mathcal{H}_x = T_x\calM$, then 
$d$ is a distance on $\mathcal{M}$, thus Theorem  \ref{thm:dd2gs} guarantees 
that $\Sigma$ is incrementally stable for any feedforward input signal ${u}(\cdot)$.

\section{Differential passivity}
\label{sec:dp}
Following the approach of \cite{Willems1972a}, we formulate
differential passivity as the 
restriction of differential dissipativity to quadratic supply rates.
To this end, we consider the external variable manifold $\calW$ as the product of 
{ an input vector space $\calU$ and an output vector space $\calY$
such that $\calU=\calY$. A consequence of working with a vector space $\calW$ is
that $T_w \calW=\calW$ for each $w\in\calW$.
In what follows, 
we will use 
$u\in\calU$ to denote the input and 
$y\in\calY$ to denote the output.}

For each $x\in\calM$, let $\bbW_x$ be a $(0,2)$-tensor field on
$\calW$ that provides an inner product on each tangent space 
$T_{w} \calW=\calW$,
denoted by $\inner{\cdot,\cdot}_{\bbW_x}$. 
For simplicity of the exposition, we write 
$\inner{\delta y,\delta u}_{\bbW_x}$ to denote
$\inner{(\delta y,0),(0,\delta u)}_{\bbW_x}$, or
$\inner{\delta y,\delta y}_{\bbW_x}$ to denote 
$\inner{(0,\delta y),(0,\delta y)}_{\bbW_x}$.

\begin{definition}
\label{def:passivity}
For each $x\in \calM$, let $\bbW_x$ be a $(0,2)$-tensor field on $\calW$.
A dynamical system $\Sigma$ 
is 
 \emph{differentially passive} if it is differentially dissipative 
with respect to a differential supply rate of the form 
\begin{equation}
\label{eq:supply_passivity}
Q(x,u,\delta u, y,\delta y) := \inner{\delta y,\delta u}_{\bbW_x}.
\end{equation}
$\Sigma$ is 
\emph{uniformly} differentially passive 
whenever $Q$ is independent on $x$. 
Finally, we say that $\Sigma$ is 
\emph{strictly} differentially passive if 
there exists a function $\alpha$ of class $\mathcal{K}$ such that 
\eqref{eq:diff_dissipativity} is restricted to 
$\dot{S} \leq -\alpha(S(x,\delta x)) + Q$.
\end{definition}

As in passivity, the next theorems show that the feedback interconnection 
of differentially passive systems is differentially passive.
\begin{theorem}
\label{thm:uniform_feedback_interconnection}
Let $\Sigma_1$ and $\Sigma_2$ be (strictly) uniformly differentially passive dynamical
systems. Suppose that $\calW_1 = \calW_2$ and that their supply rates are
based on the same $(0,2)$-tensor $\bbW$. Then, 
the dynamical system $\Sigma$ arising from the feedback interconnection
\begin{equation}
\label{eq:output_feedback_interconnection}
u_1 = -y_2 + v_1 \ , \; u_2 = y_1 + v_2, 
\end{equation}
is (strictly) uniformly differentially passive from $v=(v_1,v_2)\in \calU_1\times \calU_2$ 
to $y=(y_1,y_2)\in\calY_1\times\calY_2$.
\end{theorem}

\begin{theorem}
\label{thm:feedback_interconnection}
Let $\Sigma_1$ and $\Sigma_2$ be (strictly) differentially passive dynamical
systems. Suppose that $\calW_1 = \calW_2$ and that their
supply rates are based on the $(0,2)$-tensors $W_{x_1}$ for $x_1\in \calM_1$ 
and $W_{x_2}$ for $x_2\in\calM_2$, respectively.
Then, the dynamical system $\Sigma$ arising from the 
feedback interconnection 
\begin{equation}
\label{eq:state_feedback_interconnection}
\begin{array}{rcll}
u_1 &=& -k_2(x_2) + v_1 &  \quad k_2:\calM_2 \to \calM_1 \in C^2 \\
u_2 &=& k_1(x_1) + v_2  &  \quad k_1: \calM_1 \to \calM_2 \in C^2
\end{array}
\end{equation}
is differentially passive  from $v=(v_1,v_2)$ to $y=(y_1,y_2)$, provided that
\begin{equation}
\label{eq:interconnection_condition}
 \inner{\delta y_1, Dk_2(x_2)[\delta x_2]}_{\bbW_{x_1}}
 =   \inner{\delta y_2, Dk_1(x_1)[\delta x_1]}_{\bbW_{x_2}} 
\end{equation}
for each $x_1\in\calM_1$ and each $x_2\in\calM_2$.
\end{theorem}

The state-feedback interconnection in 
\eqref{eq:state_feedback_interconnection}
is in contrast with the classical passivity approach
that looks at systems as input/output operators. 
However, differently from
classical passivity and from uniform differential passivity,
differential passivity is an input/output characterization
of the system that depends on the \emph{
trajectories}, geometrically expressed by 
a different tensor $\bbW_x$ for each $x\in\calM$.
This lack of uniformity with respect to the solutions
of the system requires extra-effort at interconnection,
as shown by \eqref{eq:interconnection_condition}.
In this sense, the key role of the state-feedback 
\eqref{eq:state_feedback_interconnection}
is to \emph{equalize} the two tensors $\bbW_{x_1}$ and $\bbW_{x_2}$, to achieve 
the desired interconnected behavior. 
Despite the state dependence, Theorem \ref{thm:feedback_interconnection} 
can be conveniently used for design. 

\begin{example}
\label{example:input-affine}
Consider the dynamical system $\Sigma$ of equations
\begin{equation}
\label{eq:sys_input_affine}
\left\{
\begin{array}{rcl}
  \dot{x}&=& f(x) + g(x)u \\
  y &=& h(x)
\end{array} \qquad x\in \realn, \, y,u\in \real^q \ ;
\right.
\end{equation}
whose induced displacement dynamical system $\delta \Sigma$ is represented
 by \eqref{eq:sys_input_affine} and 
\begin{equation}
\label{eq:dsys_input_affine}
\left\{
\begin{array}{rcl}
  \dot{\delta x} &=& \partial_x f(x) \delta x 
  + [\partial_x g(x)u] \delta x + g(x) \delta u\\
  \delta y &=& \partial_x h(x) \delta x. 
\end{array}
\right.
\end{equation}
Let $W(x)$ a symmetric matrix for each $x\in\calM$. 
$\Sigma$ is differentially passive
with differential supply rate $\delta y^T W(x) \delta u$ if 
there exist a matrix $M(x)=\partial_{xx}m(x)$, where $m:\real^n\to\real$, 
and an invertible matrix $\Pi$ such that
\begin{equation}
 \label{eq:UC}
 \begin{array}{rcl}
    M(x)^T \partial_x[M(x)f(x)]  & \ \leq \ & 0  \\
    M(x)g(x) & \ = \ & \Pi \ ,   \\
   \partial_x h(x)^T W(x) & \ =\ & M(x)^T \Pi 
 \end{array}
\end{equation}
In fact, define $S(x,\delta x) := \frac{1}{2} \delta x^T M(x)^TM(x) \delta x$. Then, 
\begin{equation}
\begin{array}{rcl}
 \dot{S} 
& = & \delta x^T M(x) \partial_x (M(x)f(x)) \delta x \ + \\
& & + \ \delta x^T M(x)^T\partial_x (M(x)g(x)u) \delta x \ + \\
& & + \ \delta x^T M(x)^TM(x) g(x)\delta u \\
&\leq & \delta x^T M(x)^T\partial_x (\Pi u) \delta x + \delta x^T M(x)^T\Pi \delta u \\
& = & \delta x^T h(x)^T W(x)\delta u  \\
& = & \delta y^T W(x) \delta u \ .
\end{array}  \vspace{-4mm}
\end{equation}
\end{example}

\begin{figure}[htbp]
\centering
\includegraphics[width=1\columnwidth]{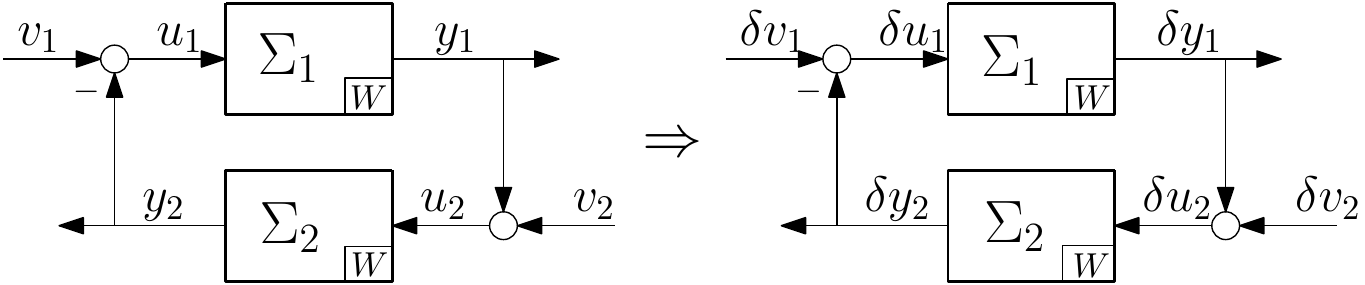} \vspace{1mm}\\
\includegraphics[width=1\columnwidth]{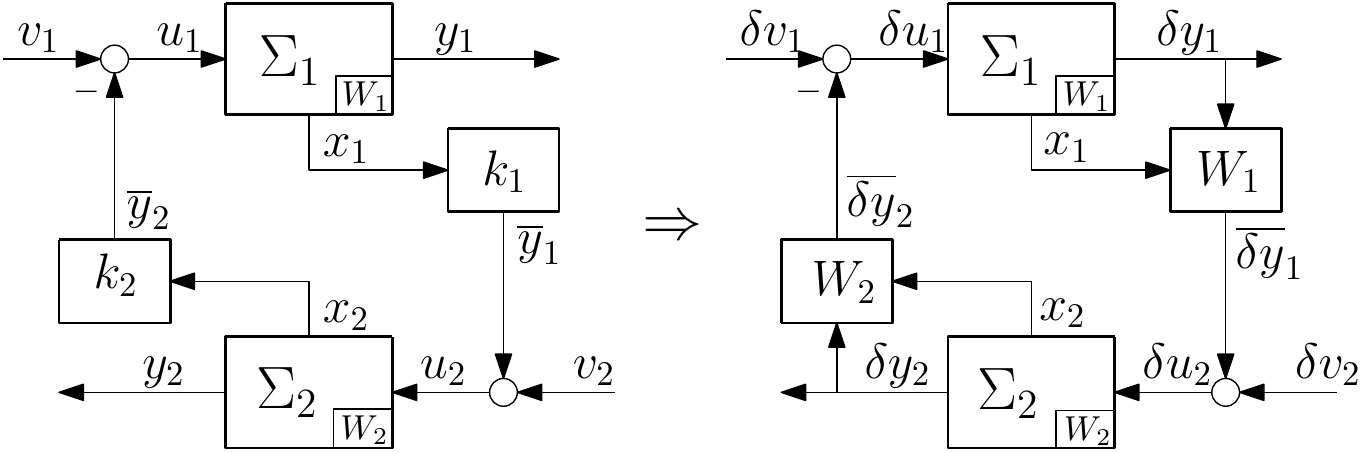} 
\caption{Interconnection of uniform differential passive systems (top).
Interconnection of differential passive systems (bottom).
$k_1(x_1)$ and $k_2(x)$
define an interconnection on \eqref{eq:sys_input_affine} 
that induces on the displacement dynamics \eqref{eq:dsys_input_affine} the cancellation 
$-\delta y_1^T W_1(x_1) W_2(x_2) \delta y_2 + \delta y_2^T W_2(x_2) W_1(x_1) \delta y_1 = 0$.
They also define new output functions $\overline{\delta y}_i$ that 
guarantee uniform differential passivity.}
\label{fig:DP} 
\end{figure}

\begin{example}
Consider the dynamical system $\Sigma$ given by
\begin{equation}
\label{eq:gen_affine_sys}
\begin{array}{rcl}
  \dot{x} &=& f(x) + g(x)u \\
        y &=& h(x) + i(x)u
\end{array} \qquad x\in \realn, \, y,u\in \real^q \ ;
\end{equation}
whose displacement dynamics is given by 
\begin{equation}
\label{eq:gen_affine_dsys}
\begin{array}{rcl}
  \dot{\delta x} &=& \partial_x f(x)\delta x + [\partial_x g(x)u] \delta x +  g(x)\delta u\\
        \delta y &=& \partial_x h(x)\delta x + [\partial_x i(x)u] \delta x + i(x)\delta u.
\end{array}
\end{equation}
Let $W(x)$ a symmetric matrix for each $x\in\calM$. 
$\Sigma$ is differentially passive
with differential supply rate $\delta y^T W(x) \delta u$ if 
there exists a matrix $M(x)=\partial_{xx}m(x)$, where $m:\real^n\to\real$,  such that  
\begin{equation}
 \label{eq:AP}
 \begin{array}{rcl}
    M(x)^T \partial_x[M(x)f(x)]  & \ \leq \ & 0  \\
  {[\partial_x h(x)]}^T W(x) &=& M(x)^T M(x)g(x)  \\
 {[}\partial_x i(x)u{]}^T W(x) &=& M(x)^T \partial_x[ M(x)g(x)u]  \\
 i(x)^T W(x) &\geq & 0. 
 \end{array}
\end{equation}
for each $x\in\realn$ and $u\in\real^d$.
In fact, using the differential storage 
$S(x,\delta x) := \frac{1}{2} \delta x^T M(x)^TM(x) \delta x$, we get
\begin{equation}
 \begin{array}{rcl}
  \dot{S} &\leq& 
 \delta x^T 
\underbrace{ M(x) [\partial_x M(x)g(x)u] }_{[\partial_x i(x)u]^TW(x)}\delta x \ + \\
& &+ \ \delta x^T \underbrace{M(x)^T M(x)g(x)}_{[\partial_xh(x)^T]W(x)}\delta u \\
&=& \delta y^T W(x)\delta u - \delta u^T i(x)^T W(x)\delta u \\ %\vspace{2mm}\\
&\leq & \delta y^T W(x)\delta u.
 \end{array}  \vspace{-4mm}
\end{equation}
\end{example}
As a final example of the section, we reconsider Example \ref{example:input-affine}
to illustrate Theorem \ref{thm:feedback_interconnection}.
{
\begin{example}
\label{example:euclidean_feedback}
Consider two systems $\Sigma_1$ and $\Sigma_2$ satisfying 
\eqref{eq:UC} respectively with matrices $M_1(x_1)=\partial_{xx}m_1(x),W_1(x_1)$ and 
$M_2(x_2)=\partial_{xx}m_2(x),W_2(x_2)$, and constant matrices $\Pi_1$ and $\Pi_2$.
The closed-loop system given by the feedback 
interconnection \eqref{eq:state_feedback_interconnection}
is differentially passive provided that 
\begin{equation}
\label{eq:DP_feedback_interconnection_cond}
\left\{\begin{array}{rcl}
 \partial_{x_2} k_2(x_2) &=& \Pi_2^T M_2(x_2) \\
 \partial_{x_1} k_1(x_1) &=& \Pi_1^T M_1(x_1)
\end{array}\right.
\end{equation}
This is an immediate consequence of 
Theorem \ref{thm:feedback_interconnection}, since 
\begin{equation}
\begin{array}{rcl}
\delta y_1^T W_1(x_1)\partial_{x_2} k_2(x_2)\delta x_2
&=&  \delta y_1^T W_1(x_1) \Pi_2^T M_2(x_2) \delta x_2 \\
&=&  \delta y_1^T W_1(x_1) W_2(x_2) \delta y_2 \\
&=&  \delta x_1^T M_1(x_1)^T\Pi_1 W_2(x_2) \delta y_2 \\
&=&  \delta x_1^T [\partial_{x_1} k_1(x_1)]^T W_2(x_2) \delta y_2 \\
\end{array}
\end{equation}
as required by \eqref{eq:interconnection_condition}.
A graphical interpretation of \eqref{eq:DP_feedback_interconnection_cond}
is provided in Figure \ref{fig:DP}.
\end{example}
}

We conclude the section by extending Theorem \ref{thm:dd2gs}.
The next theorem shows that 
a differentially passive 
dynamical system with ``excess'' of output differential passivity behaves 
like a filter: its steady-state output depends only on the signal at the input.

\begin{theorem}
\label{thm:dd2gas}
Let $\Sigma$ be a differentially passive dynamical system with
\begin{itemize}
\item differential storage $S$ such that $\mathcal{V}_x=\emptyset$ for
each $x$; 
%and gives to $\calM$ the structure of a complete metric space; 
\item differential supply rate $Q := \inner{\delta y,\delta u}_{\bbW_x} - \inner{\delta y,\delta y}_{\bbW_x} $ such that
$\inner{\delta y,\delta y}_{\bbW_x}>0$ for each $\delta y\in \calY\setminus\{0\}$ and each $x\in\calM$ (excess of output passivity).
\end{itemize}
Let $\overline{u}(\cdot):\real_{\geq 0} \to \mathcal{U}$
be a $C^2$ input signal and 
suppose that every
curve $\xi(\cdot):=(x,\overline{u},y)(\cdot)\in \Sigma$ remains bounded.

Then, for any pair $(x_1,\overline{u},y_1)(\cdot),(x_2,\overline{u},y_2)(\cdot) \in \Sigma$, 
\begin{equation}
\label{eq:asymptotic_output}
 \lim_{t\to\infty} |y_1(t) - y_2(t)| = 0 \ . \vspace{-2mm}
\end{equation}
\end{theorem}
The hypothesis of the theorem \label{thm:dd2gas} guarantees 
incremental stability of $\Sigma$ - a consequence of Theorem \ref{thm:dd2gs}.
If $\Sigma$ is strictly differentially passive, then Theorem 
\ref{thm:dd2gas} can be strengthened towards incremental asymptotic stability.
Finally, the case of $\mathcal{V}_x \neq \emptyset$ is not taken into account
here but it presents similarities with the analysis 
of Section 2.3.2 in \cite{Sepulchre1997},
about passivity with semidefinite storage functions and
stability.

\section{Examples of differentially passive electrical circuits}
\label{sec:ems}
In the first example below we show the differential passivity of a simple nonlinear RC circuit.
Differential passivity is also used in the second example
below to develop an feed-forward control strategy for an induction motor
with flux saturations. 

\begin{example}[Nonlinear RC circuit] \\
Consider the simple circuit reproduced in Figure
\ref{fig:nonlin_RC}. The nonlinearity of the circuit is due to the nonlinear relation 
$v_c = \mu(q_c)$ between
the charge $q_c$ and the voltage $v_c$ of the capacitor.
We suppose that $\mu(q_c)$ is differentiable and strictly increasing.

\begin{figure}[htbp]
\centering
\includegraphics[width=0.3\columnwidth]{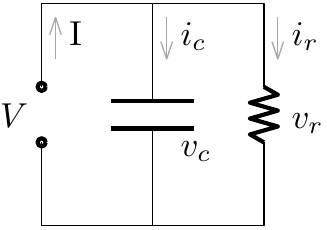}  \vspace{-2mm}
\caption{$V$,$I$ - external voltage and current. $v_c$,$i_c$ 
- capacitor voltage and current. $v_r$,$i_r$ - resistor voltage and current. }
\label{fig:nonlin_RC}
\end{figure}

The algebraic-differential description of the circuit is given by
the constitutive relations of each component and by Kirchhoff laws, 
\begin{equation}
\label{eq:nonlinRC}
\left\{
\begin{array}{rcl}
\dot{q}_c &=& i_c \\
v_c &=& \mu(q_c) \\
v_r &=& R i_r  , \, R>0
\end{array}
\right.
\ ;
\quad
\left\{
\begin{array}{rcl}
I &=& i_c + i_r  \\
 V &=& v_c \\
 v_c &=& v_r
\end{array}
\right. \ .
\end{equation}
Following \eqref{eq:ddynsys}, the displacement dynamical systems is thus represented
by  \eqref{eq:nonlinRC} and by the set of equations 
\begin{equation}
\label{eq:dnonlinRC}
\left\{
\begin{array}{rcl}
\dot{\delta q}_c &=& \delta i_c \\
\delta v_c &=& \partial_{q_c}\mu(q_c)\delta q_c \\
\delta v_r &=& R \delta i_r  
\end{array}
\right.
\ ;
\quad
\left\{
\begin{array}{rcl}
\delta I &=& \delta i_c + \delta i_r  \\
 \delta V &=& \delta v_c \\
 \delta v_c &=& \delta v_r
\end{array}
\right. \ .
\end{equation}

The circuit is differentially passive from $V$ to $I$ 
with differential storage $S(q_c,\delta q_c) := \frac{1}{2}\delta q_c^2$.
In fact, define $W(q_c) := [\partial_{q_c} \mu(q_c)]^{-1}$, then 
\begin{equation}
\begin{array}{rcl}
\dot{S} 
&=& \delta q_c \partial_{q_c} \mu(q_c) \delta q_c  \\
&=& W(q_c) \delta v_c \delta i_c  \\
&=& W(q_c) \delta V (\delta I - \delta i_r)  \\
&=& W(q_c) \delta V\delta I  - W(q_c) \delta v_r \delta i_r  \\
&\leq& W(q_c) \delta V\delta I ,
\end{array}
\end{equation}
where the last identity follows from the fact that $W(q_c) $ is greater than 0 
for each value of $q_c$, and $\delta v_r\delta i_r = R\delta i_r ^2 \geq 0$.
\end{example}

\begin{example}[Induction motor with flux saturation]\\
\label{ex:induction_motor}
We revisit the model proposed in \cite{Sullivan1996}.
The model is developed in a rotating frame at speed $\omega_s$. 
The rotor speed is denoted by $\omega_r$. Rotor and stator magnetic flux vectors
are denoted respectively by $\varphi_r$ and $\varphi_s$.
Rotor and stator currents are given by $i_r$ and $i_s$. 
The analysis below takes into account only the electrical part of the motor.
The mechanical equations are thus not detailed. Indeed,
for $\varphi_r,\varphi_s,i_r,i_s \in \mathbb{C}$, the differential
relations are given by
\begin{subequations}
\label{eq:indmot_diff}
\begin{eqnarray}
 \dot{\omega}_r &=& h(\omega_r,\varphi_r,\varphi_s,\tau_{load})  \label{eq:indmot_diff_a}\\
 \dot{\varphi}_r &=& -j \omega_g \varphi_r - R_r i_r  \label{eq:indmot_diff_b}\\
 \dot{\varphi}_s &=& - j \omega_s \varphi_s - R_s i_s + u_s  \label{eq:indmot_diff_c}
\end{eqnarray} 
\end{subequations}
where $\omega_g = \omega_s -\omega_r$, and 
$R_r$ and $R_s$ are rotor and stator resistances. $\tau_{load}$ is the (disturbance) load, and
$u_s$ is a control input. The motor 
model is completed by the algebraic relations between 
currents and fluxes, given by 
\begin{equation}
\label{eq:indmot_alg}
\left\{
\begin{array}{rcl}
i_r &=& F_r(\varphi_r) 
	+ (\frac{1}{L_r} + \frac{1}{L_l}) \varphi_r 
	- \frac{1}{L_l}\varphi_s\\
i_s &=& F_s(\varphi_s) 
	+ (\frac{1}{L_s} + \frac{1}{L_l}) \varphi_s 
	- \frac{1}{L_l}\varphi_r.\\
\end{array}
\right.
\end{equation}
$L_r$, $L_s$, and $L_l$ are the usual inductances adopted in
classical linear flux-current models, while
the nonlinear $C^2$ functions $F_r$ and $F_s$ characterize
the flux saturation. For instance, $F_r$  satisfies
a relation of the form  
$F_r(\varphi_r) = f(|\varphi_r|)\varphi_r$ where
$f$ is a monotonically increasing sector function, that is,
$f(s)\geq 0$ and $f(s)' \geq 0$, for each $s\geq 0$.
These assumptions guarantee that 
\begin{equation}
\label{eq:indmot_Fprop}
 \partial_{\varphi_r} F(\varphi_r) 
\ = \ f'(|\varphi_r|) \frac{\varphi_r\varphi_r^T}{|\varphi_r|} 
+ f(|\varphi_r|) I 
\ \geq \ 0.
\end{equation}
Indeed, the current
$i_r$ may grow faster than the flux $\varphi_r$ 
(for $F_r \neq 0$), which characterizes a limited increase of the flux 
despite large increments of the currents.
Similar assumptions hold for $F_s$.
Note that the alignment of current and flux vectors is preserved.

In what follows we will use $\Sigma$ to denote the dynamical
system represented by \eqref{eq:indmot_diff} and \eqref{eq:indmot_alg}.
Using 
$\varphi:=(\varphi_r,\varphi_s)$ and $i:=(i_r,i_s)$,
$\Sigma$ is given by the set of $C^2$ curves 
$\xi(\cdot):= (\varphi, i,\omega_r,\omega_s,u_s)(\cdot)$ that satisfy 
\eqref{eq:indmot_diff} and \eqref{eq:indmot_alg} for each $t\geq 0$.

The analysis proposed below is based on the introduction of a new
dynamical system, the \emph{virtual dynamical system} (see, for example, \cite{Wang05}), 
represented by \eqref{eq:indmot_diff_b}, \eqref{eq:indmot_diff_c} and \eqref{eq:indmot_alg},
where the relation between the rotor speed $\omega_r$ and
the flux $\varphi$ is disregarded.
To distinguish between the induction motor and the associated virtual system,
we use over-lined variables:
$\overline{\varphi}:=(\overline{\varphi}_r,\overline{\varphi}_s)$ 
and $\overline{i}:=(\overline{i}_r,\overline{i}_s)$.
Indeed, for each $\xi(\cdot) = (\varphi, i,\omega_r,\omega_s,u_s)(\cdot) \in \Sigma$, 
$\overline{\Sigma}_{\xi(\cdot)}$ is the virtual dynamical
system given by the set of curves
$(\overline{\varphi}, \overline{i},\omega_r,\omega_s,u_s)(\cdot)$
that satisfy
\eqref{eq:indmot_diff_b}, \eqref{eq:indmot_diff_c} and \eqref{eq:indmot_alg}
(expressed in the over-lined variables).

The crucial relation between $\Sigma$ and the virtual system $\overline{\Sigma}_{\xi(\cdot)}$
is that if $\xi(\cdot) \in \Sigma$, then $\xi(\cdot)\in \overline{\Sigma}_{\xi(\cdot)}$.
Exploiting this relation, it is possible to infer properties of $\Sigma$ from 
the properties of the virtual dynamical system 
$\overline{\Sigma}_{\xi(\cdot)}$.

For the virtual system $\overline{\Sigma}_{\xi(\cdot)}$, $\omega_r(\cdot)$ 
and $\omega_s(\cdot)$ are exogenous signal acting uniformly on each solution
$\overline{\varphi}(\cdot)$. Therefore
for both $\omega_s(\cdot)$ and $\omega_g(\cdot)$ 
one can consider 
$\delta \omega_g = \delta\omega_s = 0$ (see Remark \ref{rem:d_and_(2)}).
The virtual displacement dynamical system is thus given by 
\eqref{eq:indmot_diff_b}, \eqref{eq:indmot_diff_c} and \eqref{eq:indmot_alg}
(expressed in the over-lined variables) and by
\begin{equation}
\label{eq:indmot_ddiff}
\left\{
\begin{array}{rcl}
 \dot{\delta\overline{\varphi}}_r &=& 
 -j \omega_g \delta\overline{\varphi}_r 
 - R_r \delta \overline{i}_r \\
 \dot{\delta\overline{\varphi}}_s &=& 
 - j \omega_s \delta\overline{\varphi}_s 
 - R_s \delta \overline{i}_s + \delta u_s 
\end{array} 
\right. \,
\end{equation}
\begin{equation}
\label{eq:indmot_dalg}
\left\{
\begin{array}{rcl}
\delta \overline{i}_r &=& \partial F_r(\overline{\varphi}_r)\delta \overline{\varphi}_r 
	+ (\frac{1}{L_r} + \frac{1}{L_l}) \delta\overline{\varphi}_r 
	- \frac{1}{L_l}\delta\overline{\varphi}_s\\
\delta \overline{i}_s &=& \partial F_s(\overline{\varphi}_s)\delta \overline{\varphi}_s  
	+ (\frac{1}{L_s} + \frac{1}{L_l}) \delta\overline{\varphi}_s 
	- \frac{1}{L_l}\delta \overline{\varphi}_r.\\
\end{array}
\right.
\end{equation}

\eqref{eq:indmot_ddiff} and \eqref{eq:indmot_dalg} characterize respectively
a differentially passive dynamical system and a differentially passive static nonlinearity.
For  \eqref{eq:indmot_ddiff}, consider the differential storage
$V= \frac{\delta\overline{\varphi}_r^2}{2 R_r} + \frac{\delta\overline{\varphi}_s^2}{2 R_s}$.
Then, 
\begin{equation}
\label{eq:indmot_udp}
 \dot{V} 
= -\delta\overline{\varphi}_r \delta \overline{i}_r 
 - \delta\overline{\varphi}_s\delta \overline{i}_s 
+ \frac{1}{R_s} \delta\overline{\varphi}_s \delta u_s \; 
\end{equation}
which establish uniform differential passivity from $(-\delta \overline{i},\delta u_s)$ to 
$(\delta \overline{\varphi},\delta \overline{\varphi}_s)$ of the dynamical system
represented by \eqref{eq:indmot_diff_b}, \eqref{eq:indmot_diff_c} 
(expressed in the over-lined variables).

On the other hand, for \eqref{eq:indmot_dalg} we get $0 \leq \delta \overline{i}^T \delta \overline{\varphi} \ = $
\begin{equation}
\label{eq:indmot_alg_udp2}
\begin{array}{c}
=
\delta \overline{\varphi}^T 
\left( 
\underbrace{
\smallmat{\partial F_r(\overline{\varphi}_r) + \frac{1}{L_r} & 0 \\ 
0 & \partial F_r(\overline{\varphi}_s) + \frac{1}{L_s}}}_{ > 0} 
+ 
\underbrace{
\smallmat{\frac{1}{L_l} & -\frac{1}{L_l} \\ 
-\frac{1}{L_l} & \frac{1}{L_l}}}_{\geq 0} 
\right) 
\delta \overline{\varphi}.
\end{array}
\end{equation}
From \eqref{eq:indmot_udp} and \eqref{eq:indmot_alg_udp2}, the combination 
of \eqref{eq:indmot_ddiff} and \eqref{eq:indmot_dalg} guarantees
that $\overline{\Sigma}_{\xi(\cdot)}$
is strictly uniformly differentially passive from $u_s$ to $\varphi_s$,
for each $\xi(\cdot)\in \Sigma$. In fact,
\begin{equation}
 \label{eq:indmot_cloop2}
 \dot{V} \leq - \overline{\delta\varphi}^T \underbrace{M( \overline{\varphi})}_{>0}  \overline{\delta \varphi}
  + \frac{1}{R_s}  \overline{\delta \varphi}_s \delta u_s,
\end{equation}
where $M(\varphi)$ is the quantity between brackets in \eqref{eq:indmot_alg_udp2}.
Because $M(\varphi)>0$, for $\delta u_s=0$ (feedforward signal),
Theorem \ref{thm:dd2gas} guarantees that 
\begin{equation}
\label{eq:incprop}
\lim_{t\to\infty} |\overline{\varphi}_1(t) - \overline{\varphi}_2(t)| = 0
\end{equation}
for all
$(\overline{\varphi}_1, \overline{i}_1,\omega_r,\omega_s,u_s)(\cdot)$,
$(\overline{\varphi}_2, \overline{i}_2,\omega_r,\omega_s,u_s)(\cdot)$
in  $\overline{\Sigma}_{\xi(\cdot)}$ 
Note that the boundedness of these curves
is guaranteed for bounded signals $u_s(\cdot)$
by the combination of the effect of 
the dissipative terms in \eqref{eq:indmot_ddiff}
and the alignment between currents and fluxes in \eqref{eq:indmot_dalg}.

The incremental property \eqref{eq:incprop} of the virtual system 
$\overline{\Sigma}_{\xi(\cdot)}$ can be used to provide an
feedforward control design for $\Sigma$. 
For illustration purposes, in what follows we consider the goal of asymptotically regulate 
$\varphi_r$ towards a prescribed flux configuration $\varphi_r^*$. 

From \eqref{eq:incprop}, achieving the goal for the virtual system 
$\overline{\Sigma}_{\xi(\cdot)}$
is straightforward: 
if $((\varphi_r^*,\varphi_s^*), i^*,\omega_r,\omega_s,u_s)(\cdot) \in \overline{\Sigma}_{\xi(\cdot)}$
then each curve 
$(\overline{\varphi}, \overline{i},\omega_r,\omega_s,u_s)(\cdot)\in \overline{\Sigma}_{\xi(\cdot)}$
satisfies
$\lim_{t\to\infty} |\overline{\varphi}(t) - (\varphi_r^*,\varphi_s^*)(t)| = 0$.
Indeed, from \eqref{eq:indmot_diff_b}, \eqref{eq:indmot_diff_c}, 
and \eqref{eq:indmot_alg}, the feedforward input $u_s(\cdot)$ given by
\begin{equation}
\label{eq:reference}
\begin{array}{rcl}
\varphi_s^* &:=& -\frac{L_l}{R_r}\dot{\varphi_r^*} - L_l[j \omega_g 
+ (\frac{1}{L_r} + \frac{1}{L_l})]{\varphi_r^*} -  F_r({\varphi_r^*}) \\
u_s &:=& [j \omega_s + R_s (\frac{1}{L_s} + \frac{1}{L_l}) ]\varphi_s^*
  + R_s F_s(\varphi_s^*) 
	- \frac{1}{L_l}r + \dot{\varphi}_s^* 
\end{array}
\end{equation}
guarantees that 
$((\varphi_r^*,\varphi_s^*), i^*,\omega_r,\omega_s,u_s)(\cdot) \in \overline{\Sigma}_{\xi(\cdot)}$.

The reader will notice that for any given selection 
of $\xi(\cdot):= ({\varphi}, {i},\omega_r,\omega_s,u_s)(\cdot)\in \Sigma$, with $u_s(\cdot)$
given in \eqref{eq:reference}, the curve
$((\varphi_r^*,\varphi_s^*), i^*,\omega_r,\omega_s,u_s)(\cdot)$ belongs to  $
\overline{\Sigma}_{\xi(\cdot)}$.
This is a consequence
of the fact that $u_s(\cdot)$ is formulated by taking into account
explicitly $\omega_s(\cdot)$ and $\omega_g(\cdot)$. 
Thus, exploiting the fact that if $\xi(\cdot) \in \Sigma$, then 
$\xi(\cdot)\in \overline{\Sigma}_{\xi(\cdot)}$, we can conclude that
\begin{equation}
\label{eq:incprop_real}
\lim_{t\to\infty} |{\varphi}(t) - (\varphi_r^*,\varphi_s^*)(t)| = 0
\end{equation}
for all
$({\varphi}, {i},\omega_r,\omega_s,u_s)(\cdot)\in \Sigma$ with 
$u_s(\cdot)$ in \eqref{eq:reference}.
A similar (but dynamic) design
of $\overline{u}$ can be provided for the regulation of $\varphi_s$ to $\varphi_s^*$. 
\end{example}

\section{Conclusions}
\label{sec:conclusions}
{
The concept of differential dissipativity is introduced as a 
natural extension of differential stability for open systems.
The differential storage $S(x,\delta x)$ is inspired from the 
Finsler-Lyapunov function of \cite{Forni2012} and has the interpretation
of (infinitesimal) energy of a displacement $\delta x$ along a solution curve through $x$.
Extending the role of dissipativity theory for analysis and design of 
interconnections in the tangent bundle offers
a novel way to study incremental stability (or contraction) 
properties of nonlinear systems.
}

\appendix

\section{Proofs}    

\begin{proofof} \emph{Theorem \ref{thm:dd2gs}} [Sketch].
In accordance with Remark \ref{rem:d_and_(2)},
we can consider curves in $\delta \Sigma$ for $\delta u(\cdot) = 0$.
In fact, for any given pair of curves 
$(x_1,\overline{u},y_1)(\cdot),(x_2,\overline{u},y_2)(\cdot)\in\Sigma$,  
the associated parameterization satisfies
$u(\cdot,s) = \overline{u}(\cdot)$, that is,
$D u(t,s)[0,1] = 0$ for each $t$ and $s$.

As a consequence, by differential dissipativity, we have $\dot{S} \leq 0$. 
Because the differential storage $S$ is also a non-increasing
horizontal Finsler-Lyapunov function,
\eqref{eq:dd2gs} is a consequence of Theorem 3 in \cite{Forni2012}
Moreover, the case of differential storages $S$ with $\mathcal{H}_x = T_x\mathcal{M}$,
is a consequence of Theorem 1 in \cite{Forni2012}.
\end{proofof}

\begin{proofof}\emph{Theorem \ref{thm:uniform_feedback_interconnection}}
Define the differential storage $S:= S_1+S_2$ 
\footnote{When clear from the context, 
we drop the arguments of the functions to simplify the notation.}. 
The functions 
$\alpha_1$ and $\alpha_2$ below must be set to zero for the weaker property
of uniform differential passivity. 
\begin{equation}
\begin{array}{rcl}
\dot{S} 
&\leq& -\alpha_1(S_1) - \alpha_2(S_2) + \inner{\delta y_1, - \delta y_2 + \delta v_1}_{\bbW}  \\
& & + \ \inner{\delta y_2,\delta y_1+\delta v_2}_{\bbW}   \\
& = & -\alpha_1(S_1) - \alpha_2(S_2) -\inner{\delta y_1, \delta y_2}_{\bbW} 
      + \ \inner{\delta y_2,\delta y_1}_{\bbW} \\
& & + \ \inner{\delta y_1,\delta v_1}_{\bbW} 
    + \ \inner{\delta y_2,\delta v_2}_{\bbW} \\
&=& -\alpha_1(S_1) - \alpha_2(S_2) 
    + \inner{\delta y_1,\delta v_1}_{\bbW}
    + \inner{\delta y_2,\delta v_2}_{\bbW} \\
&\leq & -\overline{\alpha}(S/2) 
        + \inner{\delta y_1,\delta v_1}_{\bbW}
        + \inner{\delta y_2,\delta v_2}_{\bbW} \\
\end{array}
\end{equation}
where $\overline{\alpha}(\cdot) := \min(\alpha_1(\cdot),\alpha_2(\cdot))\in\mathcal{K}$.
In fact, 
\begin{equation}
\label{eq:sum_of_Kfunctions}
\alpha_1(S_1)+\alpha_2(S_2) 
\geq \overline{\alpha}(S_1)+\overline{\alpha}(S_2) 
\geq \overline{\alpha}\left(\frac{S_1+S_2}{2}\right)
\end{equation}
where the first inequality follows from the definition of $\overline{\alpha}$,
and the last inequality holds because
(i)~$\overline{\alpha}(S_1) \geq \overline{\alpha}\left(\frac{S_1 +S_2}{2}\right)$ for $S_1\geq S_2$;
(ii)~$\overline{\alpha}(S_2) \geq \overline{\alpha}\left(\frac{S_1 +S_2}{2}\right)$ for  $ S_2 \geq S_1$;
(iii)~$\overline{\alpha}(S_1)\geq 0$ and $\overline{\alpha}(S_2) \geq 0$.

Note that $\inner{\delta y_1,\delta v_1}_{\bbW} + \inner{\delta y_2,\delta v_2}_{\bbW} $
characterizes an inner product on the product
manifold $\calW_1\times\calW_2$.
\end{proofof}

\begin{proofof}\emph{Theorem \ref{thm:feedback_interconnection}}
Define the differential storage $S:= S_1+S_2$. As in the proof
of Theorem \ref{thm:uniform_feedback_interconnection}, 
$\alpha_1$ and $\alpha_2$ below must be set to zero for the case of differential passivity.
\begin{equation}
\begin{array}{rcl}
\dot{S} 
&\leq& -\alpha_1(S_1) - \alpha_2(S_2) \\
& &    - \inner{\delta y_1, Dk_2(x_2)[\delta x_2]}_{\bbW_{x_1}} 
       + \inner{\delta y_2, Dk_1(x_1)[\delta x_1]}_{\bbW_{x_2}} \\
& &    + \ \inner{\delta y_1,\delta v_1}_{\bbW_{x_1}}
       + \ \inner{\delta y_2,\delta v_2}_{\bbW_{x_2}} \\
&=& - \overline{\alpha}(S/2) 
    + \inner{\delta y_1,\delta v_1}_{\bbW_{x_1}}
    + \inner{\delta y_2,\delta v_2}_{\bbW_{x_2}},
\end{array}
\end{equation}
where $\overline{\alpha}(\cdot) := \min(\alpha_1(\cdot),\alpha_2(\cdot))\in\mathcal{K}$.
The last identity follows from \eqref{eq:interconnection_condition}
and from \eqref{eq:sum_of_Kfunctions}.

For each point $(x_1,x_2)$ of the product manifold $\calM_1\times\calM_2$,
$\inner{\delta y_1,\delta v_1}_{\bbW_{x_1}}
    + \inner{\delta y_2,\delta v_2}_{\bbW_{x_2}}$
defines a $(0,2)$-tensor $\bbW_{(x_1,x_2)}$ on the product
manifold $\calW_1\times\calW_2$.
\end{proofof}

\begin{proofof}\emph{Theorem \ref{thm:dd2gas}}.
Let $(x_1,\overline{u},y_1)(\cdot),(x_2,\overline{u},y_2)(\cdot) $ be any pair of 
$C^2$ curves in  $\Sigma$.
For each $s\in[0,1]$, define $u(\cdot,s) = \overline{u}(\cdot)$, and
consider a (parameterized) curve $(x,u,y)(\cdot,s):\real\to\calM\times\calW \in \Sigma$
such that $(x,u,y)(\cdot,0) = (x_1,\overline{u},y_1)(\cdot)$ and
$(x,u,y)(\cdot,1) = (x_2,\overline{u},y_2)(\cdot)$.
We assume that $(x,u,y)(\cdot,\cdot) \in C^2$.

For each $s\in[0,1]$, define
\begin{equation}
\begin{array}{rcl}
(x(\cdot,s),\delta x(\cdot,s)) &:=& (x(\cdot,s),D x(\cdot,s)[0,1]) \\
(u(\cdot,s),\delta u(\cdot,s)) &:=& (u(\cdot,s),Du(\cdot,s)[0,1]) \ = \ (\overline{u}(\cdot),0) \\
(y(\cdot,s),\delta y(\cdot,s)) &:=& (y(\cdot,s),D y(\cdot,s)[0,1]).
\end{array}
\end{equation}
Repeating the argument of Remark \ref{rem:d_and_(2)}, one can show that
$(x(\cdot,s),\delta x(\cdot,s))$ is a solution
to \eqref{eq:ddynsys} from the initial condition
$(x(0,s), \delta x(0,s))$ under the action of the input
$(\overline{u}(\cdot),0)$. Thus, the storage function $S$ satisfies
\begin{equation}
\frac{d}{dt} S(x(t,s),\delta x(t,s)) 
\leq - \inner{\delta y(t,s),\delta y(t,s)}_{\bbW_{x(t,s)}}. 
\end{equation}

By boundedness of $(x_1,\overline{u},y_1)(\cdot),(x_2,\overline{u},y_2)(\cdot) $, 
define a compact set $X\subset\calM$ such that
$x(t,s) \in X$
for each $t\geq 0$ and each $s\in [0,1]$.
$X$ depends on the range of $\overline{u}(\cdot)$, 
and on the range of parameterization of the curve
$x(0,\cdot)$.
The compactness of $X$ guarantees the existence of
a smooth $(0,2)$-tensor field $\bbW$ such that 
\begin{equation}
\label{eq:tensor_bound}
\inf_{x\in X,u\in U} \inner{\delta y,\delta y}_{\bbW_x } \geq \inner{\delta y,\delta y}_{\bbW} 
\end{equation}
for each $x\in X$. 
Then,
\begin{equation}
\frac{d}{dt} S(x(t,s),\delta x(t,s)) 
\leq - \inner{\delta y(t,s),\delta y(t,s)}_{\bbW} \ ,
\end{equation}
that is,
\begin{equation}
S(x(0,s),\delta x(0,s)) 
\geq \int_0^t \inner{\delta y(\tau,s),\delta y(\tau,s)}_{\bbW}d\tau \ .
\end{equation}
By Barbalat's lemma, for each $s\in[0,1]$,  
\begin{equation}
\label{eq:barbalat1}
\lim_{t\to\infty} \inner{\delta y(t,s),\delta y(t,s)}_{\bbW} = 0.
\end{equation}
The applicability of Barbalat's lemma follows from the fact that 
$\delta y(\cdot,s)$ is uniformly continuous for each $s\in[0,1]$. 
In fact, the range of
$(y(\cdot,s),\delta y(\cdot,s))$ is bounded for each $s\in[0,1]$.
This is a consequence of the fact that each curve to $\Sigma$
is bounded and that $\dot{S} \leq 0$.
Therefore, for each $s\in[0,1]$, 
$(y(\cdot,s),\delta y(\cdot,s))$
belongs to a compact subset of $T \calY$ that depends on
the initial condition $(x(0,s),\delta x(0,s))$ and on the input $\overline{u}(\cdot)$.
Thus, for each $s\in[0,1]$, $(y(\cdot,s),\delta y(\cdot,s))$ is uniformly continuous, since it is  a $C^1$ 
function on a compact set.

Consider now the distance $d$ on $\calY$ induced by
the Riemannian structure 
$\inner{\delta y,\delta y}_{\bbW}^{\frac{1}{2}}$. 
Then, \eqref{eq:asymptotic_output} follows from 
\eqref{eq:barbalat1} and the fact that 
\begin{equation}
d(y_1(t),y_2(t)) 
\leq \int_0^1 
\inner{\delta y(t,s),\delta y(t,\!s)}_{\bbW}^{\frac{1}{2}} ds \ .
\end{equation}
\end{proofof}

{
\begin{spacing}{1.0}

\end{spacing}
}

\end{document}